\documentclass[journal,onecolumn]{IEEEtran}

\usepackage{hyperref}
\usepackage{tabularx}
\usepackage{amsmath}
\usepackage{multirow}
\usepackage{listings}
\usepackage{graphicx}
\usepackage{booktabs}

\begin{document}

\title{DFL: High-Performance Blockchain-Based Federated Learning}

\author{\IEEEauthorblockN{Yongding Tian\IEEEauthorrefmark{1},
Zhuoran Guo\IEEEauthorrefmark{2},
Jiaxuan Zhang\IEEEauthorrefmark{3},
Zaid Al-Ars\IEEEauthorrefmark{4}}
\IEEEauthorblockA{Delft University of Technology\\
Delft, The Netherlands\\ 
Email: {y.tian-3@tudelft.nl, {Z.Guo-4, J.Zhang-42}@student.tudelft.nl \\
z.al-ars@tudelft.nl}
}}

\maketitle

\begin{abstract}

Many researchers have proposed replacing the aggregation server in federated learning with a blockchain system to improve privacy, robustness, and scalability. In this approach, clients would upload their updated models to the blockchain ledger and use a smart contract to perform model averaging. However, the significant delay and limited computational capabilities of blockchain systems make it inefficient to support machine learning applications on the blockchain.

In this paper, we propose a new public blockchain architecture called DFL, which is specially optimized for distributed federated machine learning. Our architecture inherits the merits of traditional blockchain systems while achieving low latency and low resource consumption by waiving global consensus. To evaluate the performance and robustness of our architecture, we implemented a prototype and tested it on a physical four-node network, and also developed a simulator to simulate larger networks and more complex situations. Our experiments show that the DFL architecture can reach over 90\% accuracy for non-I.I.D. datasets, even in the presence of model poisoning attacks, while ensuring that the blockchain part consumes less than 5\% of hardware resources.

\end{abstract}

\section{Introduction}

In the past ten years, people have witnessed the rise of crypto-currency and the growth of decentralized finance (DeFi). Most current blockchain applications are limited to the financial domain because it is non-sensitive to performance limitations but relies on data security and integrity. \cite{9441499} and \cite{Li2022} show some potential blockchain application areas such as Internet of Things(IoT) and health care. However, only limited practical applications exist in these areas due to the low transaction per second (TPS) and high latency of blockchain systems.

Federated learning is a machine learning (ML) training method where multiple clients send their local model updates to a central server, which will later generate a new global model and broadcast it to clients. Federated learning addresses privacy issues by keeping the training data on clients \cite{stefan2020}, but it raises trust challenges between clients and central servers. For example, clients can send incorrect models to the central server to harm the global model. Servers can also infer the training data on a specific client using the changes in its model, causing privacy leakage \cite{Protection_Against_Reconstruction_and_Its_Applications_in_Private_Federated_Learning} \cite{https://doi.org/10.48550/arxiv.2101.00159}.

To address the above challenges, some researchers suggest utilizing blockchain technology as a replacement for central servers \cite{fl_chain_1} \cite{block_fla} \cite{fl_chain_2}. This is possible because blockchain offers a transparent and honest environment for generating and broadcasting the new global model. Blockchain also offers a reliable record of the models submitted by clients, thus malicious clients can be identified and removed. Mainstream research in federated learning on blockchain uses smart contracts to perform model aggregation, which results in poor performance because generating a block usually costs over 10 seconds~\cite{chain_fl} \cite{baffle}.

This paper proposes a new blockchain architecture turning the blockchain database from a distributed ledger into a distributed proof of contribution to the ML model, where each node can generate its own blocks asynchronously, significantly improving the overall federated learning efficiency. The new architecture also shows reasonable stability and robustness, with a suitable node reputation strategy and weighted FedAvg implementation for non-I.I.D. datasets and model poising attacks.
From the ML perspective, the main contribution of this paper is an asynchronous Gossip-alike ML training method with blockchain that stores proof of contribution to the ML models. We implement this method in the DFL framework, including an executable prototype, a simulator to simulate the ML behavior, and an SDK to design reputation/model updating algorithms. This paper also mentions other algorithms, such as the reputation algorithm and weighted federated averaging algorithm, but they are not the main contributions because they are used to study the characteristics of the DFL framework.

This paper will first discuss current federated ML and blockchain systems in Section \ref{related_work}. Section \ref{high_level_overview_of_dfl} provides a high-level overview of DFL, as well as its application area and its differences with federated ML. The detailed elaboration, including data format, blockchain structure, and federated ML implementation, is presented in Section \ref{DFL_framework}. Section \ref{result_and_discussion} shows the blockchain performance and federated ML performance. The conclusions and future works are presented in Section \ref{conclusion_future_work}.
\section{Related work and background}
\label{related_work}

\subsection{Blockchain system and bifurcation}
\label{intro_blockchain_system}

Blockchain systems were first introduced in bitcoin \cite{bitcoin_paper}, a peer-to-peer, public and immutable ledger system. All transactions in bitcoin are sealed into blocks to achieve immutability. These blocks are then stored in a directed chain where each block contains the digest of the previous block. The directed chain ensures that any changes in a block will result in the change of all following blocks. Additionally, a global consensus algorithm is applied to ensure that only one chain exists through all nodes in this peer-to-peer network. Bitcoin uses proof of work (PoW) to reach the global consensus, where bitcoin miners must "dig" for a random number to make the block digest meet a certain pattern. The difficulty of finding such a random number varies with the computation power in the whole blockchain network - it becomes more difficult to find a qualified random number with more computational power in total. 


In a blockchain system, the computation power of the network is measured by the speed of generating blocks, or equivalently, the interval between two adjacent blocks. This interval time is important for the blockchain system as it determines the probabilities of bifurcation. The term bifurcation describes a situation where in a blockchain network, two nodes generate two correct blocks simultaneously. Since each node only accepts the first block received, one of the blocks would be rejected by over half of all miners. All miners will try to generate the next block based on their current received block. In bifurcation, one block is more likely to generate the next block once it is accepted by most miners because more miners mean more computation resources to find the next block. So as time goes by, one chain will be much longer than the other chains and be accepted by all miners, where the global consensus is reached again. The bifurcation can significantly degrade the performance of the blockchain system and should be avoided. One standard solution is increasing the mining difficulty, which leads to a longer block generation interval. Meanwhile, a longer interval also brings higher latency. However, The bifurcation problem exists in blockchain systems no matter the consensus algorithms.


\subsection{Federated machine learning (federated ML)}

Federated ML continues to draw increasing attention from the research community. A typical federated ML system contains a central server and many clients. The central server is responsible for aggregating and updating the models trained by clients, which are then sent back to clients for the next training iterations. This method of aggregating and updating the central model is called FedAvg \cite{fedavg}. Federate ML has advantages in both data privacy since the training data is privately kept to the client, and model training quality achieves high accuracy for I.I.D., non-I.I.D. and unbalanced datasets \cite{fed_ml}.

However, federated ML also brings new vulnerabilities to the system, including data poisoning \cite{federated_ml_data_poison_1} \cite{federated_ml_data_poison_2}, model poisoning \cite{federated_ml_model_poison_1} and privacy attack from the server \cite{federated_ml_server_privacy_attack_1}. Data poisoning and model poisoning aim to provide wrong models to the central server to harm the training process of the global model. Besides, the central server is able to recover the protected dataset stored on the client by applying Generative Adversarial Networks (GANs) \cite{Zaid_fl_attack} \cite{federated_ml_server_privacy_attack_1}.

Even though federated ML decentralizes ML systems from central servers to client devices, the central server must be kept to perform model aggregation and updating. This makes the central server a single point of failure to the entire system. To deal with this single point of failure and potential attacks on privacy, solutions were proposed where the central server can be substituted by a blockchain \cite{block_fl} \cite{chain_fl} \cite{baffle}. 

\subsection{Federated ML on blockchain}

Bitcoin, as the first blockchain system, has limited functionality as it only provides a distributed and immutable ledger. The next-generation blockchain systems, such as Ethereum \cite{wood2014ethereum}, provide smart contracts to allow users to execute specific small applications on the blockchain system. The smart contract is more reliable because its source code is open to the public on the blockchain and all miners will execute it to ensure the integrity of the results. Some researchers hence attempt to integrate federated ML with Ethereum and replace the model aggregating server with smart contracts \cite{chain_fl} \cite{baffle}. However, applying smart contracts on public blockchain systems is expensive due to the miner fee. Federated ML on blockchains would also be slow because every data aggregation and computation will be performed on all nodes. A brief cost analysis is available in \cite{baffle}, where federated ML costs over $10^8$ Ethereum gas - a cost that would be even greater for larger and more complex ML tasks. This is also the reason why \cite{chain_fl} uses a private blockchain system to avoid high costs. In addition to the above problems, the storage also becomes expensive for federated ML because every model from all ML clients will be stored on all blockchain nodes.

Based on the previous discussion, here is a summary of the problems of current ML approaches to blockchain systems.

\begin{itemize}
  \item Low blockchain performance: limited transactions per block and high latency resulting from long block generation interval.
  \item Mining cost: ML applications must afford the miner fee, which is an extra expense for ML applications.
  \item Excessive storage cost: ML models will be stored on all blockchain miner nodes because the models are stored in a transaction, which will be packed into a block. All miners keep the blockchain database so every single model will be stored on each miner's disk.
\end{itemize}

The above three problems hinder the integration of federated ML on the blockchain, even though there are already many theoretical works to improve the performance of federated ML on the blockchain. For example, \cite{1_bit_awarded_federated_distillation} proposes a compression method to reduce network communication during training; \cite{8843900} focuses on privacy issues in industrial IoT applications. However, only a few of them are built on a real blockchain system \cite{chain_fl} \cite{baffle}.

In this paper, we propose a practical framework (DFL) with 1.~a deployable executable, 2.~a simulator to simulate ML performance, and 3.~an SDK to support further research.

\section{High-level overview of DFL}
\label{high_level_overview_of_dfl}



DFL is a peer-to-peer (p2p) network where users collaborate for model training by exchanging model parameters. The models are updated asynchronously through i) training when sufficient local data become available to the nodes and ii) averaging when a given number of models are received. DFL applies a local ledger to register the training contribution of each node to be used as an incentive mechanism. DFL implements partial consensus which is more effective in this case, because partial consensus allows each node to do model averaging when they receive models rather than waiting for a block to be finalized. DFL introduces a distributed reputation system to mitigate dataset and model attacks as partial consensus makes attackers hiding easier compared with traditional blockchain-based FL solutions. 

Section \ref{partial_consensus} talks about the partial consensus and its prototype - gossip algorithm in ML. Section \ref{reputation} elaborates on the distributed reputation system. Section \ref{possible_applications} and Section \ref{comparision_of_dfl} show the potential applications of DFL and compare DFL with existing related projects.

\subsection{Gossip algorithm and partial consensus}
\label{partial_consensus}

Gossip algorithms have been a popular choice for decentralized federated learning research, as they enable the sharing of models in peer-to-peer (P2P) networks \cite{dfl_gossip} \cite{gossip_algorithm} \cite{Braintorrent}. DFL also utilizes a gossip mechanism, in which each node only shares its models with a limited number of neighbors and invites them to be witnesses of its contribution to the ML models. While gossip algorithms offer a decentralized approach to model sharing, they may not provide the same level of transparency and trust as a blockchain-based system because single node doesn't obtain the information of the whole network. 

From a blockchain perspective, the gossip mechanism used in DFL is called "partial consensus", as the ML model (transaction) is only broadcast to a limited range of nodes, rather than the entire blockchain network as in global consensus. Partial consensus is suitable for ML applications for two main reasons. Firstly, the order in which transactions arrive is not critical, as a transaction that arrives late can still be used in the current averaging round. Secondly, the model weights contained in the transactions are not critical. Two models with the same accuracy, as determined by the same test dataset, may have different model weights.

As mentioned in Section \ref{intro_blockchain_system}, global consensus in blockchain systems can lead to bifurcation, and many contemporary blockchain systems increase the block interval to avoid this issue. In contrast, the partial consensus in DFL relaxes the strict synchronization requirement and thus allows for improved model communication efficiency and reduced training time. The role of blockchain databases also shifts from an information-sharing system to proof of contribution for the ML models.
Each node generates and maintains its own blockchain database, which records feedback from its neighbors (i.e., witnesses). However, partial consensus can also introduce new vulnerabilities, such as the potential for malicious behaviors like model poisoning to go unnoticed in the network. To address these concerns, a reputation system will be introduced in the next section to incentivize honest participation and mitigate the risk of malicious behavior.



\subsection{Distributed reputation}
\label{reputation}

There are already some existing reputation systems in the federated ML area. For example, \cite{reputation_incentive_mechanism} proposed using reputation as an incentive mechanism. The reputation system in DFL is responsible for distinguishing the malicious node in partial consensus. A malicious node can be a node frequently broadcasting low-accuracy models, or performing a data poisoning attack, or performing a model poisoning attack. The reputation is distributed and different nodes may rate the same node independently. For example, considering a three-node system - the reputation of node C in node A might be different from that stored in node B. 

The reputation can be updated according to any rational ML model criterion. In this paper, we mainly consider updating the reputation according to the model's accuracy. More specifically, each node will have its own local reputation record of its neighbors and will update the reputation record based on the accuracy of neighbors' models calculated by itself based on its local dataset. Broadcasting models with higher accuracy earn the node a higher reputation in the system. The reputation will be stored locally and not shared with other nodes. 

To protect the system, a malicious node should have a lower reputation in the reputation system. Meanwhile, the influence of malicious models should be lowered as well. We propose using the weighted federated averaging method \cite{weighted_fedavg}, a modified version of FedAvg, in this paper. The modified FedAvg mechanism works together with the reputation system to keep the federated ML secured and stable.

In Section \ref{result_and_discussion}, we implemented a sample reputation and weighted average system and obtained promising results. However, there are many other factors that can affect the implementation of this system, such as the malicious rate, training performance, and data distribution. In future work, we plan to investigate the impact of these factors on the reputation and weighted FedAvg system and develop more universal implementations that can handle a wide range of variations.

\subsection{Possible applications}
\label{possible_applications}

This subsection talks about potential applications for DFL. These applications are only feasible with proper reputation algorithms and federated averaging algorithms, which are not covered in this paper.

One potential application is replacing current blockchain-based federated ML systems for less training time and lower resource consumption. Meanwhile, the blockchain database can serve as an incentive mechanism and allow further reward distribution, avoiding the miner fee issue in traditional blockchain-based federated ML systems.

Another possible application is allowing independent developers to deploy their ML models without running expensive cloud servers or paying miner fees. For example, they can embed DFL in their apps to collect data and train models. The reward can be replaced with in-app benefits to compensate users for hardware resources.

DFL also makes it possible to build community-driven ML ecosystems, in which developers can propose ML models that are trained using data contributed by users and advertise them to attract miners to collaborate on the training process. Smart contracts on traditional blockchain systems can be used to receive DFL blockchain data and distribute token rewards to miners as an incentive for their participation. These tokens can then be used for ML inference in the future if the model is adequately trained, and people can buy the tokens from miners to pay the inference fee. As a result, ML models that are more useful and have higher accuracy are expected to have higher token values on the traditional blockchain, reflecting their perceived value in the ecosystem.

\subsection{Comparison}
\label{comparision_of_dfl}
\cite{9441499} summarizes existing blockchain-based federated ML systems and identifies four major issues that are addressed by these systems: (1) single-point failure, (2) poison attack, (3) lack of motivation, and (4) privacy leakage. Table \ref{tab:comparison_with_other_blockchain_based_FML_systems} lists the addressed issues by each system analyzed in \cite{9441499} and makes a comparison with DFL. The reasons for each addressed issue in DFL are listed below:
\begin{itemize}
    
    \item Single point failure: there is no central server or central ML model, so there is no single point of failure.
    
    \item Privacy leakage: the model in each transaction is a mixture of model training from the dataset and model updating with transactions from other nodes. In addition, each transaction will only be broadcast to limited other nodes due to the $ttl$ mechanism. These two features make privacy attacks much harder.
    
    \item Mining fee: the DFL network avoids mining fees by allowing each node to generate its own blockchain data, unlike traditional blockchain-based federated ML systems that rely on blockchain systems and require mining fees for executing smart contracts.
    
    \item Synchronicity in the blockchain: DFL is an asynchronous system because it does not require global consensus among the nodes in the network. This asynchronicity can reduce the strict synchronization requirements of traditional blockchain systems to allow for faster training of the ML models.
    \item Lack of motivation: the blockchain data on that node proves the node's contribution to ML models. The contribution is quantified as improvements in accuracy measured by neighbors. The final reward should be calculated by an algorithm (not covered in this paper) involving the blockchain data and the reputations of that node.
    
    \item Poison attack: while the design of a reputation algorithm to prevent poison attacks is not covered in this paper, DFL provides an SDK for further research on this topic. As shown in Section \ref{result_and_discussion}, dataset poison attacks are less effective in DFL because most nodes are innocent. Model poison attacks require additional research.
\end{itemize}

\begin{table*}[h]
\caption{Comparison with other blockchain-based federated ML systems}
\label{tab:comparison_with_other_blockchain_based_FML_systems}
\resizebox{\textwidth}{!}{%

\begin{tabular}{llll}
\hline
System & Addressed issues & Type of blockchain & Consensus mechanism \\ \hline
BlockFL\cite{block_fl} & Single point failure, lack of motivation & Public & PoW \\
FLChain\cite{fl_chain_1} & Single point failure, lack of motivation & Public & – \\
FLChain\cite{fl_chain_2} & Single point failure & Public & PBFT/PoW \\
RFL\cite{reputation_incentive_mechanism} & Poison attack, lack of motivation & Consortium & PBFT \\
DeepChain\cite{deep_chain} & Single point failure, lack of motivation, privacy leakage & Public & Blockwise-BA \\
FedBC\cite{fed_bc} & Single point failure, privacy leakage & – & – \\
BC-FL\cite{bc_fl} & Single point failure, lack of motivation & Public & Pow \\
BlockFLA\cite{block_fla} & Single point failure, poison attack, lack of motivation & Public \& private & PoW \& PBFT \\
Chain FL\cite{chain_fl} & Single point failure & Private & PoA \\
PSFL\cite{psfl} & Poison attack & – & – \\
BFEL\cite{bfel} & Single point failure, poison attack & Public \& consortium & DPoS/PBFT \& PoV \\
Biscotti\cite{biscotti} & Single point failure, poison attack, privacy leakage & Public & PoF \\
VFChain\cite{vf_chain} & Single point failure, poison attack & Public & Algorand \\
BLADE-FL\cite{blade_fl} & Single point failure & Public & PoW \\
BFLC\cite{bflc} & Single point failure, poison attack, lack of motivation & Public & Committee \\
VBFL\cite{vbfl} & Single point failure, poison attack, lack of motivation & Public & PoS \\ \hline
DFL & \begin{tabular}[c]{@{}l@{}}Single point failure, privacy leakage,\\ mining fee, synchronicity in blockchain,\\ lack of motivation (with a proper rewarding algorithm),\\ poison attack (with a proper reputation mechanism)\end{tabular} & Public & Partial consensus \\ \hline
\end{tabular}
}
\end{table*}


There are several projects that are similar to DFL but are not based on blockchain technology. For example, IPLS \cite{IPLS} is a decentralized federated learning system that uses the inter-planetary file system (IPFS) for model dissemination. In terms of federated learning, IPLS applies the segmented model to save bandwidth, which is also different from DFL. Another project, BrainTorrent \cite{Braintorrent}, is a peer-to-peer federated learning system that is similar to DFL in terms of the federated learning process. However, BrainTorrent requires nodes to actively choose their peers and collect models, whereas DFL uses a broadcast mechanism for model dissemination. Overall, these projects demonstrate that there are alternative approaches to federated learning that can be implemented without the use of blockchain technology.

\section{Technical description of DFL}
\label{DFL_framework}

\begin{figure*}
    \centering
    \includegraphics[width=0.8\textwidth]{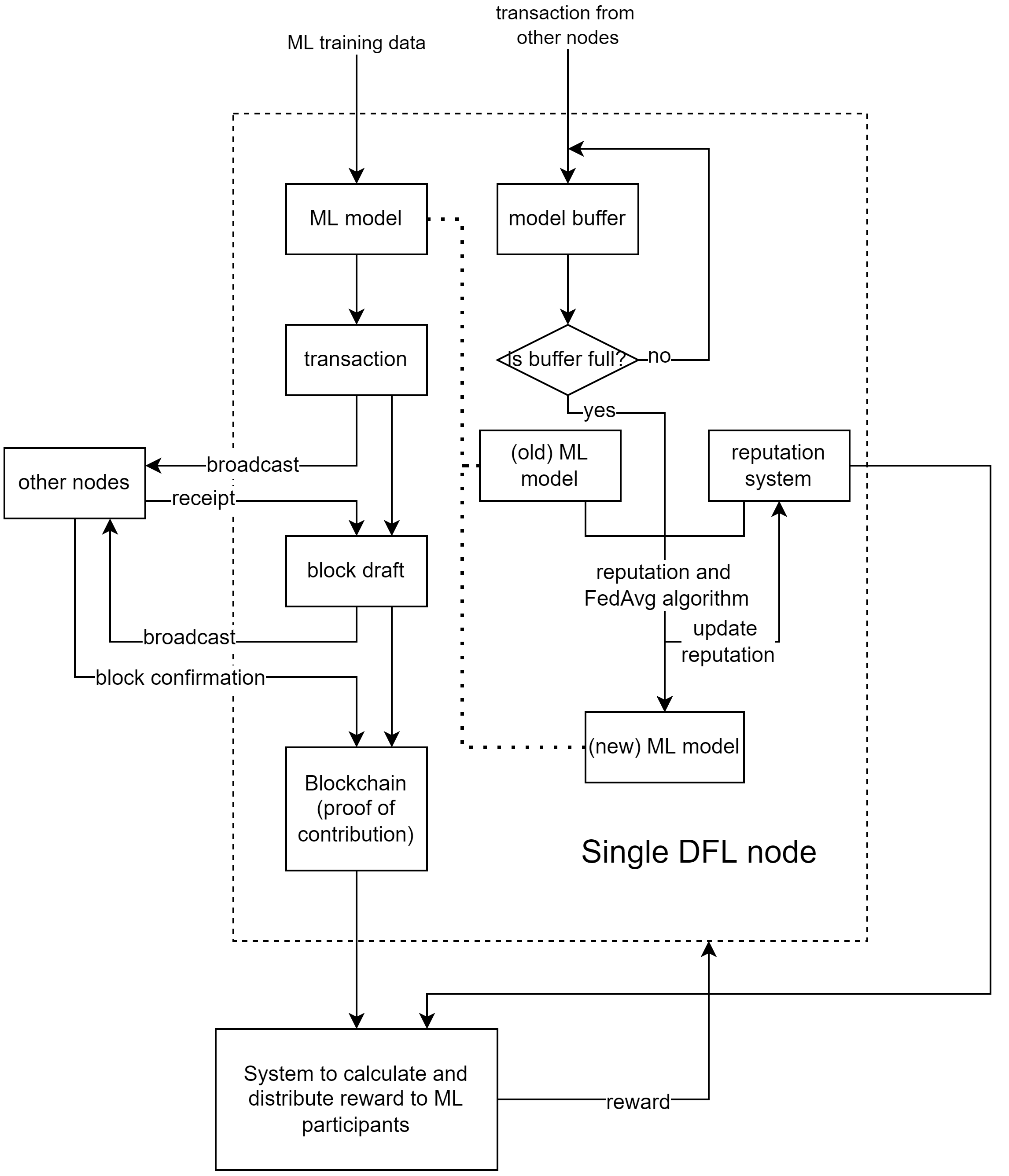}
    \caption{DFL overview}
    \label{DFL_overview}
\end{figure*}

Figure \ref{DFL_overview} shows a high-level overview of the DFL workflow. The workflow consists of two event-driven phases. The left phase is triggered when the node has collected enough "ML training data" for training (e.g. when the training data size becomes equal to/larger than the ML batch size). The node will create a "transaction" that includes the new "ML model", and broadcast it to other nodes. After several rounds of communications, the node can generate the "blockchain (proof of contribution)". The right phase is triggered when the node has received enough transactions (controlled by "model buffer" size) from other nodes to update the ML model and ends by updating the "(new) ML model" and "reputation system". 

The blockchain data serves as proof of contribution and can be submitted to a dedicated reward calculation system to quantify the contributions to ML models and distribute the rewards to the DFL nodes. This could be an incentive mechanism to encourage each DFL node to train and share their models.

\subsection{DFL workflow}
\label{DFL_workflow}

In this section, a fully-connected four-node DFL system will be used as an example to explain the workflows and structures. Please keep in mind that DFL does not rely on a fully-connected network to operate, less p2p connections also ensure the system's correctness.

Table \ref{notion_table} lists some important concepts which will be used in this section. In the table, either one of the mainstream asymmetric encryption methods, such as ECDSA and RSA, could be applied in the system.

\begin{table*}
    \centering
    \caption{Notion table for Section \ref{DFL_workflow}}
    \label{notion_table}
    \begin{tabularx}{\textwidth}{l|X}
    \toprule
    Notion & Definition\\
    \midrule
    $pri\_key_{node}$ & The private key for a node.\\
    $pub\_key_{node}$ & The public key for a node.\\
    $address_{node}$ & An unique identifier for a node.\\
    $hash(content)$ & The hash for certain content. The content will be immutable once $hash$ is calculated.\\
    $signature(pri\_key, hash)$ & The signature generated by the $pri\_key$ and $hash$.\\
    \midrule
    $generator$ & The node of generating transactions and blocks.\\
    $neighbors\ of\ x$ & All nodes whose network distance to node x is 1.\\
    $further\ neighbors\ of\ x$ & All nodes whose network distance to node x is larger than 1.\\
    \midrule
    \end{tabularx}
\end{table*}

\subsubsection{Node setup}

Each node in the DFL network generates a pair of public/private keys and an address to join the DFL network. The key pair can be generated by most cryptology methods, while the address is produced according to the rule $address_{node} = hash(pub\_key_{node})$.

\subsubsection{Generating transaction and receipt}

\begin{figure*}
    \centering
    \includegraphics[width=0.5\textwidth]{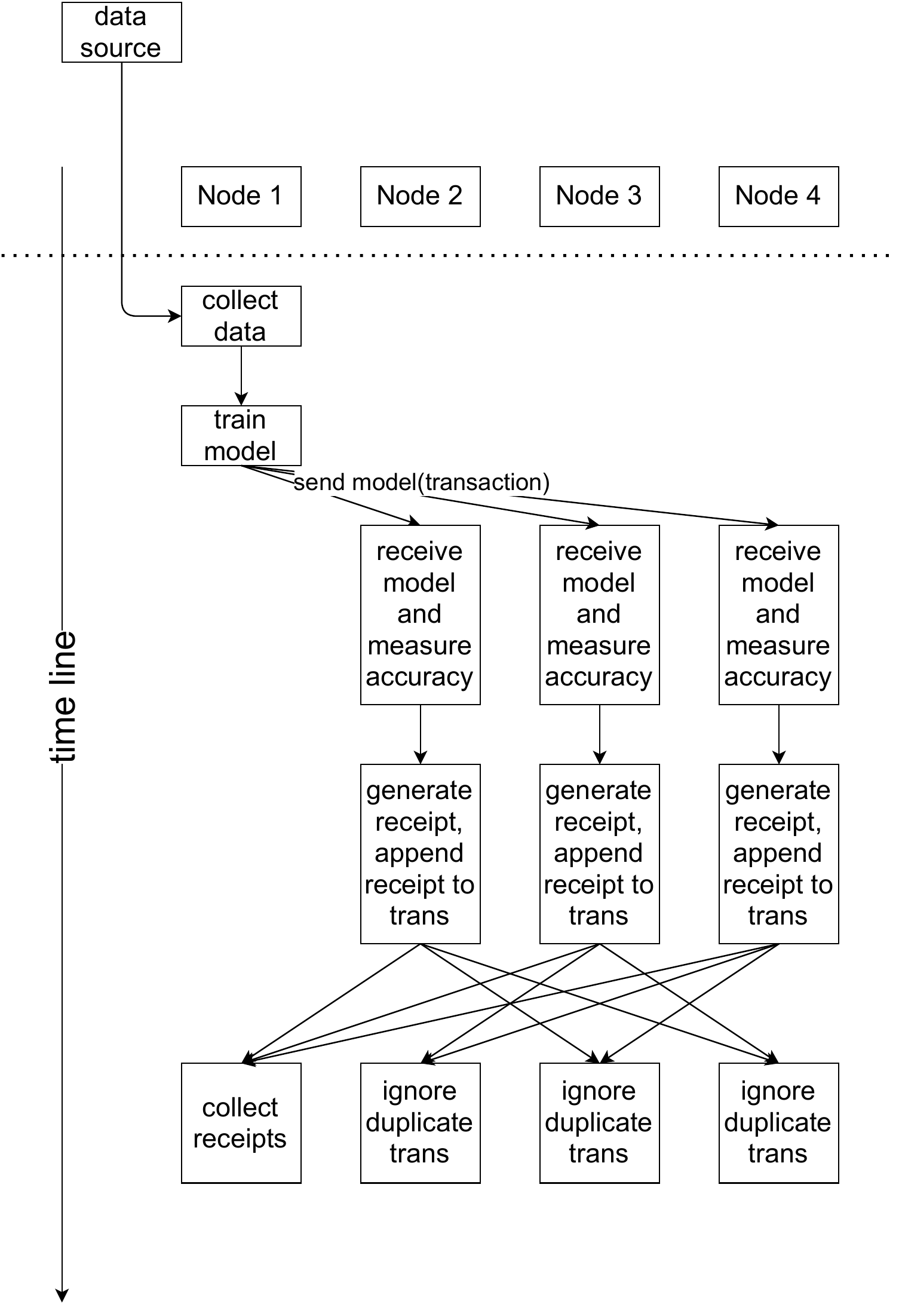}
    \caption{Procedures of generating transaction and receipt }
    \label{transaction generation}
\end{figure*}

Figure \ref{transaction generation} describes the procedures to generate transactions.

Suppose in our DFL system, node 1 is training at the beginning. The model weights after training are sent to the neighbors of node 1, i.e. nodes 2, 3, and 4 since the network is fully connected. Once other nodes receive the weights, they will start to measure the model accuracy based on their own collected dataset. Each receiver will generate a receipt describing the accuracy of the model, which is then appended at the end of the transaction. The transaction, together with the receipt, will be sent from each receiver to all its neighbors. For instance, node 2 as receiver of model weights from node 1, will send the transaction and receipt to, not only node 1 but also nodes 3 and 4 as well. However, nodes 3 and 4 will ignore the transaction and receipt from node 2, as this transaction has been processed beforehand. Node 1 will collect the receipt for generating the block afterward.

\subsubsection{Generate block}

\begin{figure*}
    \centering
    \includegraphics[width=0.5\textwidth]{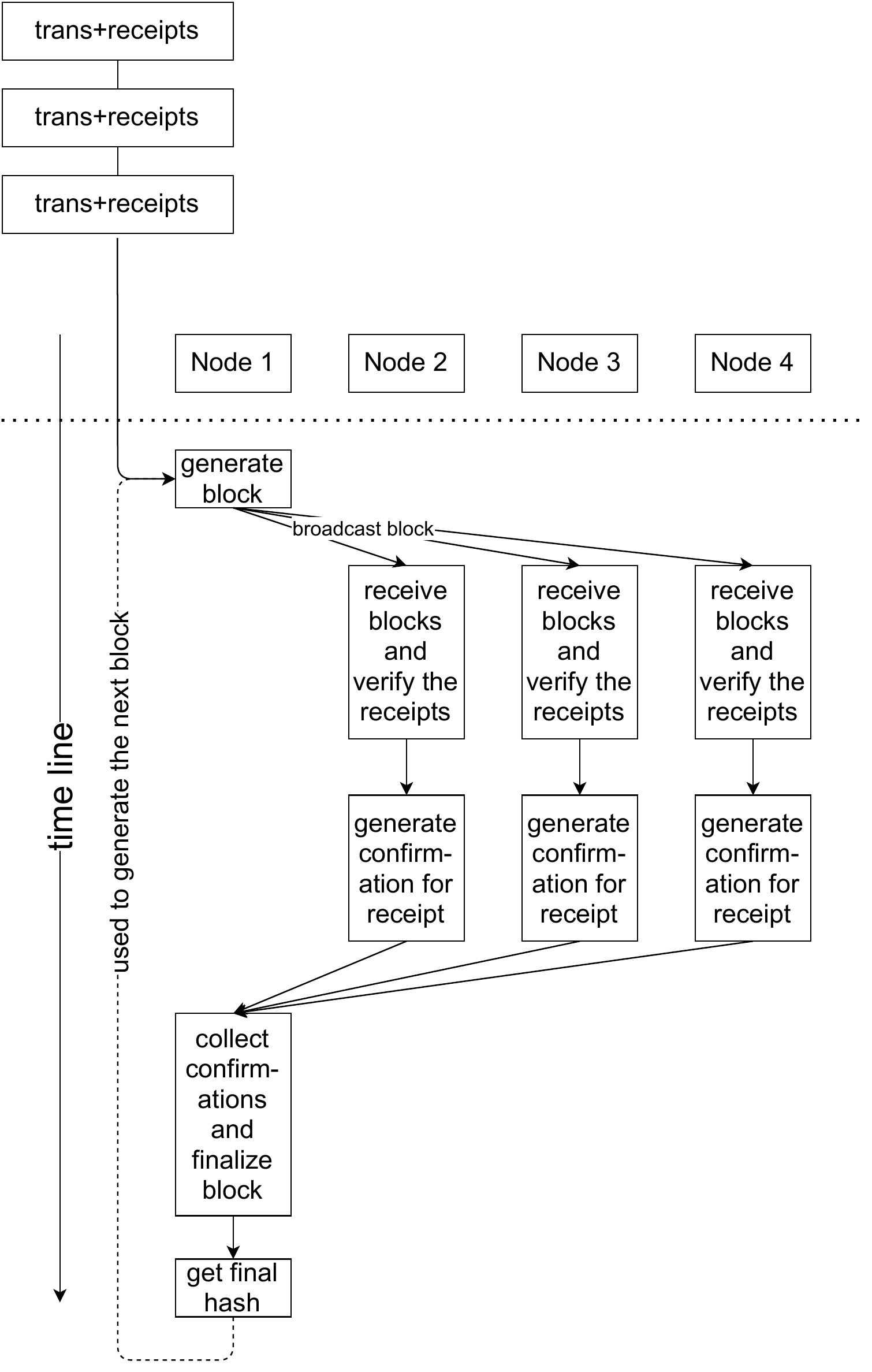}
    \caption{Procedures of generating a block}
    \label{generate block}
\end{figure*}

Figure \ref{generate block} describes the procedures to generate blocks. Node 1 will gather a certain amount of transactions and corresponding receipts after a certain time. A block would be generated to record the transactions and receipts in order to make them immutable.

The procedure for generating such a block can be divided into two phases. The block generator creates a draft block in the first phase. The second phase is to gather confirmations from neighbors of the block generator and  finalize a block. With confirmations appended to the final block, the block generator cannot easily modify the previous block as it cannot fake a confirmation.


Retake the four-node system as an example. Node 1 sends a draft block to node 2, then node 2 will look through the receipts to figure out whether some receipts are produced by node 2 and generate confirmations if there are any. Node 1 will gather all confirmations from nodes 2, 3, and 4 and append the confirmations to finalize the block. The digest of the final block will then be used to generate the next block.

\begin{figure}
    \centering
    \includegraphics[width=0.8\columnwidth]{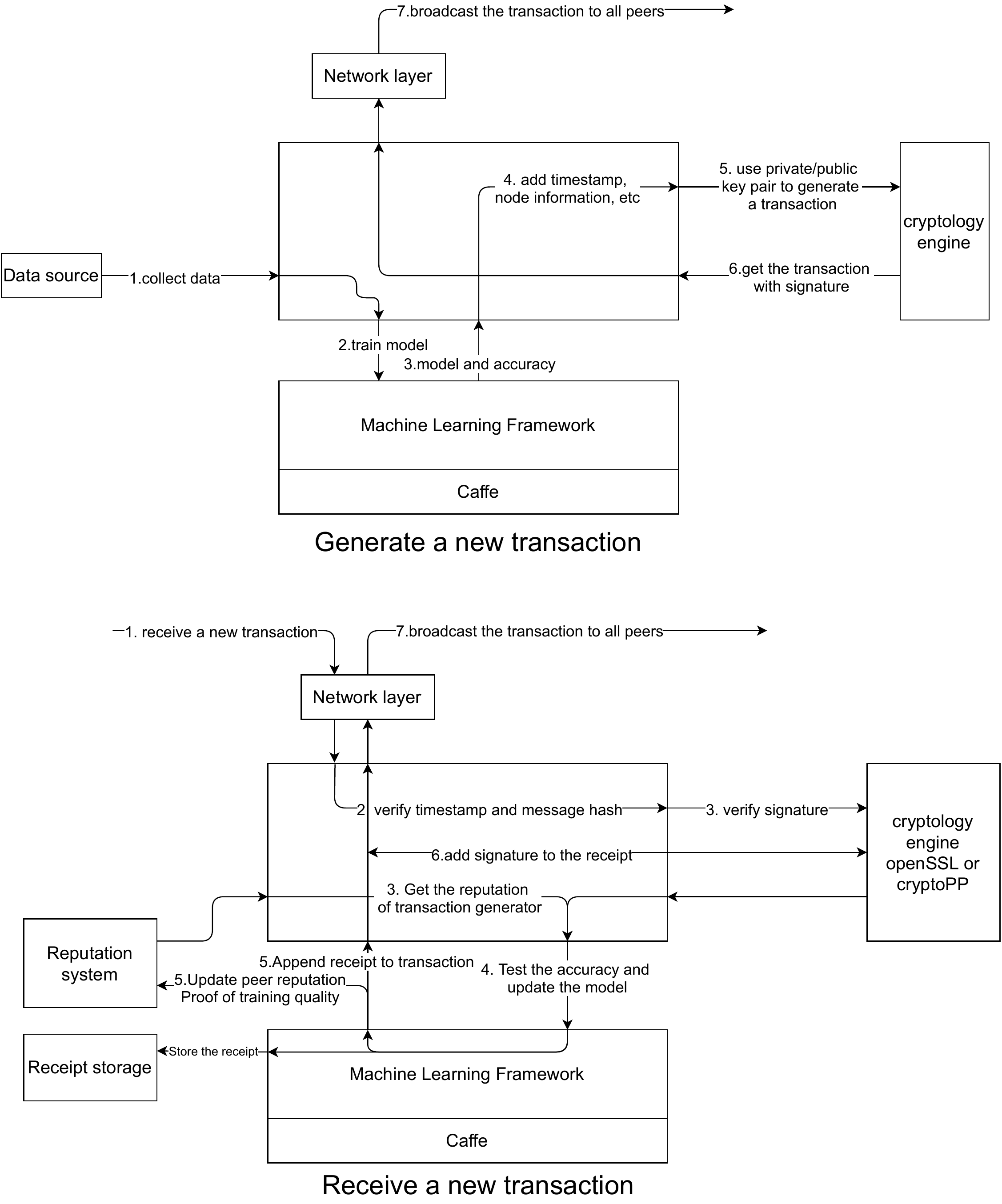}
    \caption{Generating transaction (block diagram)}
    \label{generate_transaction_block_diagram}
\end{figure}

\begin{figure}
    \centering
    \includegraphics[width=0.8\columnwidth]{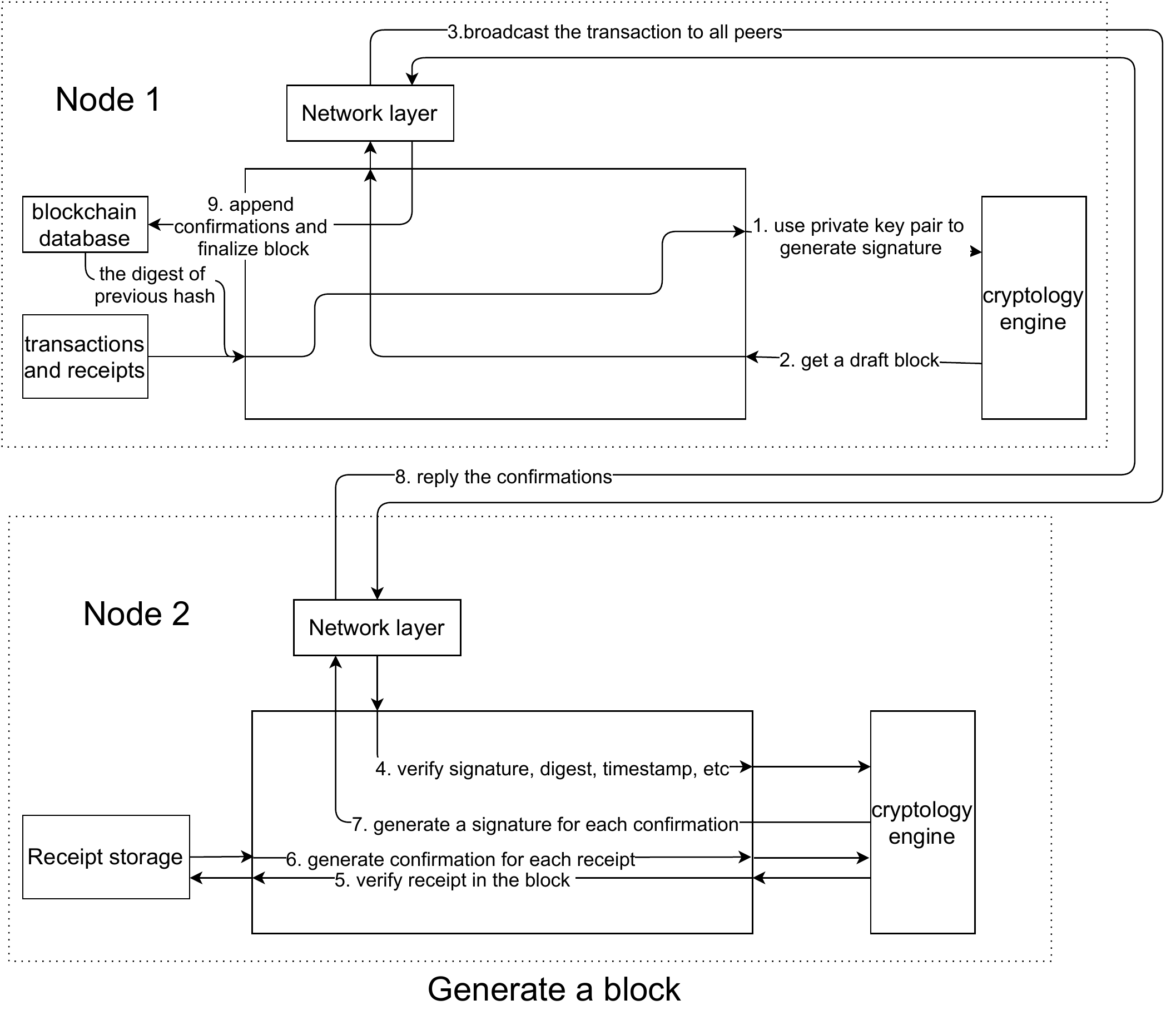}
    \caption{Generating block (block diagram)}
    \label{generate_block_block_diagram}
\end{figure}

The component diagram is available in Figure \ref{generate_transaction_block_diagram} and Figure \ref{generate_block_block_diagram}. The number before each step indicates the order. These steps present the same process flow as Figure \ref{transaction generation} and Figure \ref{generate block}, so they will not be discussed again.

\begin{figure}
    \centering
    \includegraphics[width=0.9\columnwidth]{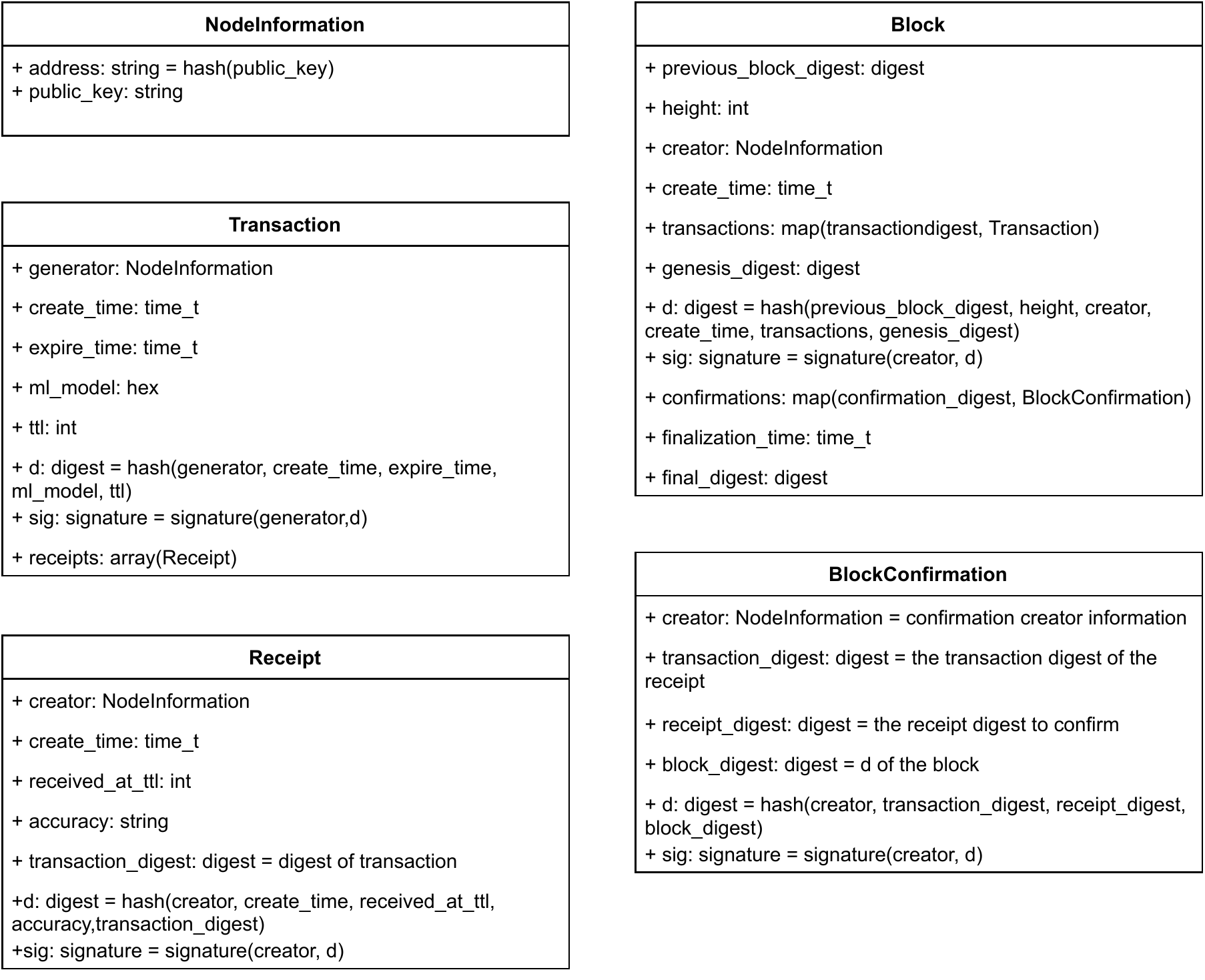}
    \caption{UML graph for DFL blockchain system}
    \label{uml:dfl}
\end{figure}

\subsection{DFL blockchain data format (UML available in Figure \ref{uml:dfl})}
\label{DFL_blockchain_data_format}

In this section, the data formats in the blockchain system will be discussed based on the workflow in previous sections. To be specific, the data formats include the format of the transaction, the transaction receipt, the block, and the block confirmation. These data formats will be used to prove the robustness of the DFL blockchain system.

These data formats will be illustrated in Table \ref{blockchain_notion_table} with some special notions.

\begin{table*}
    \caption{Notion table for Section \ref{DFL_blockchain_data_format}}
    \label{blockchain_notion_table}
    \centering
    \begin{tabularx}{\textwidth}{X|X}
    \toprule
    Notion & Definition\\
    \midrule
    $=\ (operation)$ & This data is obtained by performing the operation.\\
    $hash\ (content)$ & The hash for a certain content, the content will be immutable once $hash$ is calculated.\\
    $signature\ (pri\_key, hash)$ & The signature generated by the $pri\_key$ and $hash$.\\
    $digest\ protected\ (digest\ d)$ & This data is protected by the digest d. If the original data is modified, the digest verification will fail.\\
    \textit{signature protected (signature s, node o)} & This data is definitely generated by $node\ o$ and has not been modified once the signature verification passes. Because a signature is generated based on a digest, all signatures-protected data is also digest-protected.\\
    \midrule

    \end{tabularx}
\end{table*}

\subsubsection{Node information}


The name of a node is its address. Since the node address is decoupled from the network address, it is hard to access a specific node by its name. This property could achieve similar anonymity as the blockchain since nodes can only receive a trusted transaction from the node but hard to access the generator. For neighbors, they can know each other's network address because they will receive a transaction with no receipt.

\subsubsection{Transaction format}


Among all the transaction fields, the $generator$, $create\_time$, $expire\_time$, $ml\_model$, and $ttl$ are protected by the signature, which can only be modified by the transaction creator. A modified transaction with a changed digest will be treated as a new transaction.

The $ttl$ (time to live) is an integer value decreased by 1 when the transaction is forwarded by a node. The transaction will cease being forwarded if its $ttl$ reaches zero. The $ttl$ value is critical for the range of partial consensus. Larger $ttl$ results in a wider partial consensus range with more nodes involved in the accuracy testing for this transaction. So an infinite $ttl$ should be avoided because it might consume too much computing resource. The mechanism to punish a node generating large-$ttl$ transactions can be achieved in the reputation system.

The $expire\_time$ is designed to avoid a late arrived transaction, whose ML model has already been outdated, i.e. processed by other nodes. Those outdated ML models can harm the overall training process and reduce the creator's reputation. This expiration time can as well avoid attackers' attempts to decrease the reputation of the victims by delaying their transactions.

\subsubsection{Receipt format}


The $creator$ is the receipt creator, corresponding to nodes 2,3, and 4 in the four nodes example in Figure \ref{transaction generation}. The $transaction\_digest$ is the digest of the received transaction. The transaction digest will not change after appending new receipts because the transaction digest does not take $receipts$ into the computation. The $received\_at\_ttl$ is calculated by the following equation.

\begin{equation}
    \begin{aligned}
    received\_at\_ttl = min(trans.ttl, trans.receipts.received\_at\_ttl) - 1
    \end{aligned}
\end{equation}

The $accuracy$ is tested based on the receipt creator's dataset. In DFL, all nodes will not shuffle their dataset, so a model can have different accuracy for different nodes, especially for non-I.I.D. dataset. The accuracy will be forwarded back to the transaction creator and other nodes. This accuracy information might be helpful for other nodes because it assists to judge the receipt creator's reputation, but it has not yet been implemented in the current DFL version.

\subsubsection{Block format}


The $transactions$ in the block are the final collected transactions which contain receipts. Neighbors should verify the correctness of the receipts and generate a block confirmation after verification. The block generator should collect a certain number of confirmations from other nodes in the network in order to finalize the block.
The $genesis\_digest$ is the digest of the genesis block, which records the ML network structure to ensure all nodes use the same ML network. So all nodes in DFL should have the same $genesis\_digest$ if they are training the same model.
The $final\_digest$ is calculated after the block generator collects enough confirmations for each transaction, and will be chained to the next block.




\subsection{DFL blockchain properties}

\subsubsection{Asynchronous}
One of the key features of DFL is its ability to operate asynchronously, as it does not rely on global consensus. This means that the generating transaction, block, receipt, and confirmation processes on different nodes can be performed independently of each other. Additionally, the ML process and the generating blockchain process can also be asynchronous, as long as it is not necessary to record the contribution in the current or next block. This asynchronous nature of DFL allows nodes to operate independently, which can enhance the scalability and efficiency of the system.

\subsubsection{Anonymity}
All nodes are anonymous on the DFL network, as their IP addresses are hidden behind their node addresses and their data packets are only available to their direct peers. While this anonymity provides some level of protection for nodes, it is important to note that the node address itself may serve as an identifier for an attacker who wants to recover the dataset held by a victim. There are already several blockchain projects, such as Monero \cite{monero}, that aim to enhance node anonymity. However, this paper will not further discuss the issue of anonymizing node addresses.

\subsubsection{Immutability}
In DFL, blocks become immutable once they are finalized. This is because neighbors will not generate block confirmations for a receipt more than once, and the block generator cannot modify any content in the receipts or confirmations, as they are protected by the signatures of other nodes. The immutability of blocks in DFL helps to ensure that the contribution to the ML models cannot be modified. This is important for maintaining the integrity of the system and ensuring that all nodes are properly incentivized for their participation.

\subsection{Discussion}
Here are some discussions and comments about the DFL system:

\begin{itemize}
    \item The DFL blockchain data can be deleted to save storage after the contributors have received their rewards. Suppose the reward system for DFL is a smart contract on a traditional blockchain. It is important to minimize the amount of data that is stored on the traditional blockchain, so we should keep the on-chain data minimized to reward algorithm arguments, such as model accuracies. 
    \item In DFL, the communication cost depends on the ML model size, $ttl$, and the number of peers. In traditional blockchain-based federated ML, the communication cost depends on the model size (or gradient size) and the communications in the blockchain system. So the practical communication cost is application-specific and difficult to compare.
    \item Server-side attacks are not applicable because there are no centralized servers. Peer-side attacks, such as model poisoning and dataset poisoning, are a potential concern in DFL.
    \item Privacy attacks are harder to perform in DFL than in a federated learning system because the ML models will only be shared with a limited number of nodes in the network. The ML models in the blockchain data can also be deleted before submitting to the rewarding system if the rewarding algorithm does not require the ML models.
    \item As for network failures, failure to broadcast transactions can cause some transactions to be missed, resulting in fewer contributions to the network and fewer blocks. Failure to send draft blocks results in fewer blockchain confirmations. In practical DFL deployment, each node only needs to collect a certain percentage of confirmations in order to finalize a block. Disconnected nodes can catch up with the rest of the network after re-connection by using a model updating mechanism that gives higher weights to more accurate machine learning models.
    \item An attack method to fake contributions is possible in DFL. This attack method can be described as an attacker that steals the model from other legitimate nodes and generates a new transaction that includes the copied model to gain contributions. Verifying the digest of a transaction does not help because the attacker can slightly modify the model weights to get a completely different digest while retaining similar accuracy. There are two potential ways to defend against this attack. One is to use Trusted Execution Environments (TEEs) \cite{blockchain_TEE} on all nodes to protect the model data. The other way is to use homomorphic encryption \cite{cloud_homomorphic_encryption} to encrypt the model, which allows other nodes to use the model without decrypting it.
    
\end{itemize}

\section{Result and discussion}
\label{result_and_discussion}

This section uses experiments to measure the DFL performance and simulations to investigate the ML properties regarding malicious nodes and non-I.I.D. datasets \footnote{DFL source code available: \href{https://github.com/twoentartian/DFL/tree/9d8d0a2a4e5e05e91460d005a567a8c73b608739}{\texttt{https://github.com/twoentartian/DFL/tree/9d8d0a2a4e5e05e91460d005a567a8c73b608739}}}. 
We also upload the simulation configuration files and reputation binary files to make the simulation results re-producible. 
The ML model used in our experiment and simulations is LeNet \cite{LeNet}, and the dataset is MNIST \cite{deng2012mnist}. We use the same hyperparameters as the LeNet example in Caffe \cite{lenet_caffe_hyperparameter} because improving the ML models is not our focus. We implement two types of reputation and weighted FedAvg methods. The first implementation is a slightly modified version of federated averaging without reputation. The following formula describes the model updating procedure on each node:

\begin{equation}
\label{eq:HalfFedAvg}
    model_{next} = \frac{\sum_{n=1}^{N} \frac{1}{N} \cdot model_{n} + model_{prev}}{2}
\end{equation}

The $model_{next}$ is the output model of this model updating iteration, $model_{n}$ represents the received models from other nodes, $model_{prev}$ is the local output model of the last model updating iteration. $N$ represents the total number of models in the model buffer. This formula does not utilize the reputation mechanism, and we call this formula "\texttt{HalfFedAvg}" in this paper because, in this formula, each node's own model contributes to 50 percent of the new model, while models received from other nodes together contribute the other 50 percent.

The second implementation is called "\texttt{reputation-0.05}", as described in the following formula:

\begin{equation}
    weight = reputation \cdot accuracy
\end{equation}

\begin{equation}
\label{eq:Reputation_0.05}
    model_{next} = \frac{\sum_{n=1}^{N} \frac{weight_{n}}{\sum_{m=1}^{N}weight_m} \cdot model_{n} + model_{prev}}{2}
\end{equation}

$reputation$ is the reputation of the model generator, ranging from 0 to 1. The node with the lowest accuracy will be punished by reducing its reputation by 0.05 in each iteration (the punishment will stop if the reputation is already 0). $accuracy$ stands for the calculated accuracy from the model in the transaction, ranging from 0 to 1. Suppose there are $N$ models in the FedAvg buffer, the total weight is obtained by summing all weights (like in the denominator $weight_m$), and the previous model is marked as $model_{prev}$. A special case in "\texttt{reputation-0.05}" is that the reputations of all peers become 0. In this case, the mechanism will be replaced by "\texttt{HalfFedAvg}".



We note here that the algorithms and mechanisms introduced above are not the main contributions of this paper. The main contribution of this paper is the DFL framework itself, and the above algorithms are used to show the DFL framework is possible to be applied in malicious and non-I.I.D. cases.

\subsection{DFL performance (experiment)}
The goal of this experiment is to investigate the performance of the DFL network. The built-in performance profiler in the DFL executable can measure the CPU time consumed by each task in the DFL components (generate block, generate transactions, update models and receive transactions). The GPU acceleration in Caffe is disabled to ensure the metric of CPU time is fair for ML and blockchain tasks.
We provide two DFL testnets, a four-node net, and a two-node net. Both testnets use the "\texttt{reputation-0.05}" mechanism.

\subsubsection{The two-node net}
The two-node test net is configured as each node will receive eight randomly-selected data samples from the MNIST training dataset per second. The training batch size is 64, so each node will perform a training every 8 seconds in theory. A transaction will be generated and broadcast immediately after the training. Each node will generate a block after broadcasting 4 transactions, equivalent to 32 seconds. The test batch size is 100, and each node will randomly choose 100 samples from all received samples. We do not use the test dataset here because choosing from received samples is closer to practical cases. On a higher level, other DFL nodes with different received samples can replace the role of the test dataset. The FedAvg buffer size is set to 4, causing the FedAvg and reputation mechanism to run roughly every 32 seconds.

\begin{figure}[h]
    \centering
     \includegraphics[width=0.5\textwidth]{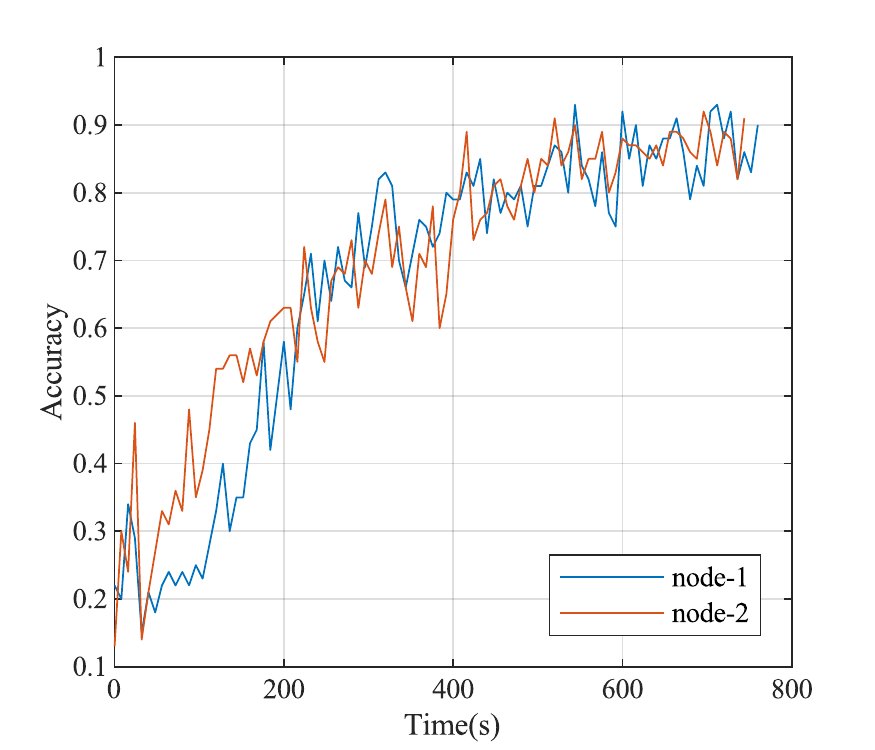}
    \caption{Accuracy graph for 2 nodes}
    \label{fig:experiment_2_node_accuracy}
\end{figure}

Figure \ref{fig:experiment_2_node_accuracy} shows the accuracy graph. The accuracy is tested based on local received data on each node. It causes a rapid increase (from 0s to 30s) in accuracy at the beginning of training. The rapid drop near 32 second is caused by the FedAvg process, which merges the models from other nodes with different datasets. Table \ref{tab:2_node_performance} shows the performance result and Table \ref{tab:2_node_blockchain_statistics} shows the blockchain statistics. The performance result indicates that the blockchain part (blockchain overhead) only consumes a small portion of CPU time, the network (broadcast transaction, gather confirmation and broadcast generated transaction) consumes much more, and ML (calculate accuracy, calculate self-accuracy and measure accuracy) costs the most. In DFL, the blockchain data is proof of contribution to ML models, and spending less time generating such proof is always desirable.

Notice that the time in Table \ref{tab:2_node_performance} is CPU time while the time in Figure \ref{fig:experiment_2_node_accuracy} (x-axis) is wall clock time. Because DFL is a multi-threading application, summing all time in Table \ref{tab:2_node_performance} will result in a larger value than wall clock time.

\begin{table}[h]
\centering
\caption{Two node performance result, unit: CPU second}

\begin{tabular}{l|cc|ccc}
\hline
\multirow{6}{*}{node-1} & \multicolumn{2}{c|}{Generate block (s)} & \multicolumn{3}{c}{Generate transaction (s)} \\ \cline{2-6} 
 & Gather confirmation & \begin{tabular}[c]{@{}c@{}}Blockchain \\ overhead\end{tabular} & Broadcast transaction & Measure accuracy & \begin{tabular}[c]{@{}c@{}}Blockchain \\ overhead\end{tabular} \\ 
 & 21.52 & 3.48 & 59.10 & 362.83 & 18.31 \\ \cline{2-6} 
 & \multicolumn{2}{c|}{Update model (s)} & \multicolumn{3}{c}{Receive transactions (s)} \\ \cline{2-6} 
 & Calculate self-accuracy & \begin{tabular}[c]{@{}c@{}}Blockchain \\ overhead\end{tabular} & \begin{tabular}[c]{@{}c@{}}Broadcast generated \\ transaction\end{tabular} & Calculate accuracy & \begin{tabular}[c]{@{}c@{}}Blockchain \\ overhead\end{tabular} \\ 
 & 82.94 & 0.94 & 80.57 & 341.73 & 5.50 \\ \hline
\multirow{6}{*}{node-2} & \multicolumn{2}{c|}{Generate block (s)} & \multicolumn{3}{c}{Generate transaction (s)} \\ \cline{2-6} 
 & Gather confirmation & \begin{tabular}[c]{@{}c@{}}Blockchain \\ overhead\end{tabular} & Broadcast transaction & Measure accuracy & \begin{tabular}[c]{@{}c@{}}Blockchain \\ overhead\end{tabular} \\ 
 & 22.22 & 2.40 & 72.81 & 312.30 & 7.18 \\ \cline{2-6} 
 & \multicolumn{2}{c|}{Update model (s)} & \multicolumn{3}{c}{Receive transactions (s)} \\ \cline{2-6} 
 & Calculate self-accuracy & \begin{tabular}[c]{@{}c@{}}Blockchain \\ overhead\end{tabular} & \begin{tabular}[c]{@{}c@{}}Broadcast generated \\ transaction\end{tabular} & Calculate accuracy & \begin{tabular}[c]{@{}c@{}}Blockchain \\ overhead\end{tabular} \\ 
 & 77.16 & 0.43 & 44.89 & 310.63 & 3.62 \\ \hline
\end{tabular}%

\label{tab:2_node_performance}
\end{table}

\begin{table*}[h]
\centering
\caption{Two node blockchain statistics}

\begin{tabular}{ccccc}
\hline
 & Transaction/Block & Confirmation/Block & Peers & Block \\ \hline
node-1 & 4 & 4 & 1 & 24 \\ \hline
node-2 & 4 & 4 & 1 & 24 \\ \hline
\end{tabular}%

\label{tab:2_node_blockchain_statistics}
\end{table*}

\subsubsection{The four-node net}
\begin{figure}[h]
    \centering
     \includegraphics[width=0.5\textwidth]{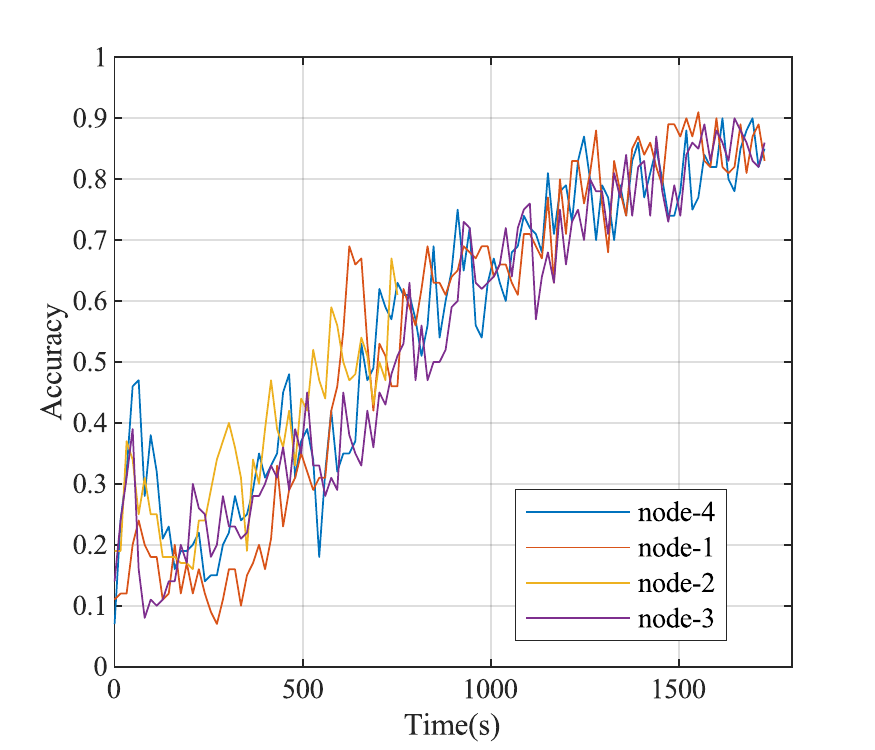}
    \caption{Accuracy graph for 4 nodes}
    \label{fig:experiment_4_node_accuracy}
\end{figure}

Some configuration modifications are applied to the four-node DFL network: the training sample received by each node is decreased from 8 to 4 per second; the FedAvg buffer size is increased from 4 to 8; the transaction count in each block is increased from 4 to 12. The network topology is fully-connect. The results are shown in Figure \ref{fig:experiment_4_node_accuracy}, Table \ref{tab:4_node_performance}, and Table \ref{tab:4_node_blockchain_statistics}.

\begin{table*}[h]
\centering
\caption{Four node performance result, unit: CPU second}

\begin{tabular}{l|cc|ccc}
\hline
\multirow{6}{*}{node-1} & \multicolumn{2}{c|}{Generate block (s)} & \multicolumn{3}{c}{Generate transaction (s)} \\ \cline{2-6} 
 & Gather confirmation & \begin{tabular}[c]{@{}c@{}}Blockchain \\ overhead\end{tabular} & Broadcast transaction & Measure accuracy & \begin{tabular}[c]{@{}c@{}}Blockchain \\ overhead\end{tabular} \\ 
 & 56.63 & 3.17 & 71.72 & 364.72 & 8.28 \\ \cline{2-6} 
 & \multicolumn{2}{c|}{Update model (s)} & \multicolumn{3}{c}{Receive transactions (s)} \\ \cline{2-6} 
 & Calculate self-accuracy & \begin{tabular}[c]{@{}c@{}}Blockchain \\ overhead\end{tabular} & \begin{tabular}[c]{@{}c@{}}Broadcast generated \\ transaction\end{tabular} & Calculate accuracy & \begin{tabular}[c]{@{}c@{}}Blockchain \\ overhead\end{tabular} \\ 
 & 90.34 & 0.74 & 214.48 & 1944.49 & 23.33 \\ \hline
\multirow{6}{*}{node-2} & \multicolumn{2}{c|}{Generate block (s)} & \multicolumn{3}{c}{Generate transaction (s)} \\ \cline{2-6} 
 & Gather confirmation & \begin{tabular}[c]{@{}c@{}}Blockchain \\ overhead\end{tabular} & Broadcast transaction & Measure accuracy & \begin{tabular}[c]{@{}c@{}}Blockchain \\ overhead\end{tabular} \\ 
 & 105.66 & 4.00 & 88.89 & 399.76 & 8.86 \\ \cline{2-6} 
 & \multicolumn{2}{c|}{Update model (s)} & \multicolumn{3}{c}{Receive transactions (s)} \\ \cline{2-6} 
 & Calculate self-accuracy & \begin{tabular}[c]{@{}c@{}}Blockchain \\ overhead\end{tabular} & \begin{tabular}[c]{@{}c@{}}Broadcast generated \\ transaction\end{tabular} & Calculate accuracy & \begin{tabular}[c]{@{}c@{}}Blockchain \\ overhead\end{tabular} \\ 
 & 99.67 & 3.45 & 299.28 & 1438.01 & 20.07 \\ \hline
\end{tabular}%

\label{tab:4_node_performance}
\end{table*}

\begin{table}[h]
\centering
\caption{Four node blockchain statistics}

\begin{tabular}{ccccc}
\hline
 & Transaction/Block & Confirmation/Block & Peers & Block \\ \hline
node-1 & 12 & 35.11 & 3 & 9 \\ \hline
node-2 & 12 & 35.67 & 3 & 9 \\ \hline
\end{tabular}%

\label{tab:4_node_blockchain_statistics}
\end{table}

In theory, the \texttt{Confirmation/Block} value should be 36 because each transaction has 3 confirmations, and there are 12 transactions in a block. The real value is slightly smaller because some confirmations are not collected in time. In practical deployment, the rule of finalizing a block could be "collecting at least 80 percent of all confirmations" or "collecting at least 60 percent of confirmations for each transaction" to make trade-offs on the blockchain robustness and performance.

As shown in Table \ref{tab:4_node_performance}, most CPU resources are consumed by calculating the accuracy of received models, and this time consumption is directly proportional to the number of direct neighbors. Given that the number of neighbors is restricted to a specific maximum and that the workflow on each node is asynchronous from that of other nodes, DFL does not suffer from blockchain-related scalability issues as the number of nodes in the network increases.

To illustrate that DFL has higher performance compared to other blockchain-based FL systems, we compare it to Baffle \cite{baffle}, a federated learning system based on a private Ethereum network with Proof-Of-Authority consensus. The results of this comparison are shown in Table \ref{tab:compare_dfl_baffle}, which indicates that the time consumed by the DFL blockchain part is 10 times smaller than Baffle. It is important to note here that the blockchain performance in Baffle is affected by the "chunk size" parameter (which is not an issue for DFL). Additionally, the reason why we do not use wall-clock time for DFL in Table \ref{tab:compare_dfl_baffle} is that it is determined by the manually-set training data injection rate.

\begin{table}[h]
\centering
\caption{Time comparison: DFL vs Baffel}
\begin{tabular}{|l|l|l|}
\hline
 & DFL & BAFFLE \\ \hline
Model & LeNet & 2-layer DNN, 500 perceptrons each layer \\ \hline
Number of nodes & 2$\sim$4 & 16$\sim$128 \\ \hline
ML-related activities & 2000s$\sim$2500s (CPU time) & 70s$\sim$90s (wallclock time) \\ \hline
Blockchain-related activities & $\sim$200s (CPU time) & \begin{tabular}[c]{@{}l@{}}2000s (chunk size \footnote{chunk size is a blockchain parameter in BAFFLE \cite{baffle}} 16kB) - \\ 11000s (chunk size 2kB) (wallclock time)\end{tabular} \\ \hline
\end{tabular}
\label{tab:compare_dfl_baffle}
\end{table}

Despite the conclusion that the blockchain part only consumes little CPU resources, the result shows another conclusion that the CPU time consumed by calculating accuracy for models from other nodes is proportional to the number of peers. In our experiments, calculating accuracy is a bottleneck for performance because it delays replying to receipts. Re-performing the experiment with GPU accelerators may solve this bottleneck. In real deployments, the number of peers should be controlled to save hardware resources.

\subsection{ML properties (simulation)}
\label{subsection:ml_propertires}
In this section, an investigation is conducted to assess the effect of non-I.I.D. datasets and malicious nodes on machine learning. Due to some limitations in deploying larger DFL networks, we use a simulator to replace the actual deployment.

\begin{itemize}
    \item The DFL simulator removes the blockchain part.
    \item In DFL deployments, each node is asynchronous because they run on different computers. The simulator is a single process, and its behavior is synchronous, so we use a virtual time scale, called "Tick", to simulate the asynchronous behavior. For example, each node receives a batch of training data every "x" ticks, the "x" is a random value between 8 to 12.
    \item The DFL simulator does not simulate delay. The delay includes network delay and the time spent on ML. For example, a node will know the accuracy of a model from other nodes immediately once broadcasted. 
    \item In simulator, the accuracy values are always calculated based on the MNIST test dataset rather than based on the actual received data samples received by each node.
\end{itemize}

We investigate three non-I.I.D. types of dataset. The first type is that each node only gets a dataset with two specific labels. For example, node-0 only gets training data labeled with 0 and 1; node-1 only gets training data labeled with 1 and 2. 
For the other two types of non-IID samples, we sampled a probability distribution over the class labels for each node based on a symmetric Dirichlet prior. In particular, we investigated the cases \texttt{alpha}=1 and \texttt{alpha}=0.1 where \texttt{alpha} is a concentration parameter of the Dirichlet distribution.
Smaller \texttt{alpha} results in a more non-I.I.D. dataset. For non-I.I.D. cases described by the Dirichlet distribution, we will regenerate the probabilities of each label and repeat the simulation 10 times to avoid extreme results.

There are two types of malicious nodes: dataset-poisoning attackers and model-poisoning attackers. Dataset attackers always use random samples (all pixels are uniform random numbers from 0 to 1) to train models and send them to other nodes. Model attackers train the model normally but always send randomly generated models to other nodes. The weights of the model are uniform random numbers from 0 to 0.001. We choose 0.001 as the maximum weight value to avoid NaN (floating-number overflow) error in Caffe. The output of the random model is also random because of the last Softmax layer.

The DFL simulation network contains 10 nodes, and each node actively connects to two other peers, so in theory, each node will have 4 peers if no overlapping occurs. The network topology is randomly generated by a DFL tool, and all simulations in this section use the same network topology. The FedAvg buffer size is 4 regardless of peer count. The training batch size is 64, and the test batch size is 100. Each node will get a batch of training data at the training tick, which is determined as the current training tick plus a random number ranging from 8 to 12. The transaction $ttl$ (defined in Figure \ref{uml:dfl}) is fixed to 1. 

The DFL simulator provides two metrics. The first metric is the model accuracy of each node, and the second metric is a value to describe how different the models are. The following formula calculates the value:

\begin{equation}
    \textit{difference} = \frac{\sum_{n=1}^{N-1} {| model_n - model_{n-1} |} + | model_0 - model_n | }{N}
\end{equation}

The $N$ represents the number of nodes \footnote{also represents the number of models because each node has one model}, $model_n$ indicates the sum of the weights for a specific layer of model $n$. $model_n$ is a vector of  4 elements if the ML model has 4 layers. The operator $|\cdot|$ calculates the absolute value of the vector elements. The output $difference$ is a vector where each element represents the difference of a layer among $N$ models.

As mentioned previously, there are two reputation and FedAvg mechanisms and both of them will be used in the simulation. The first mechanism is "\texttt{HalfFedAvg}" described in Equation \ref{eq:HalfFedAvg} and "\texttt{Reputation-0.05}" described in Equation \ref{eq:Reputation_0.05}.

\begin{figure}
    \centering
     \includegraphics[width=\textwidth]{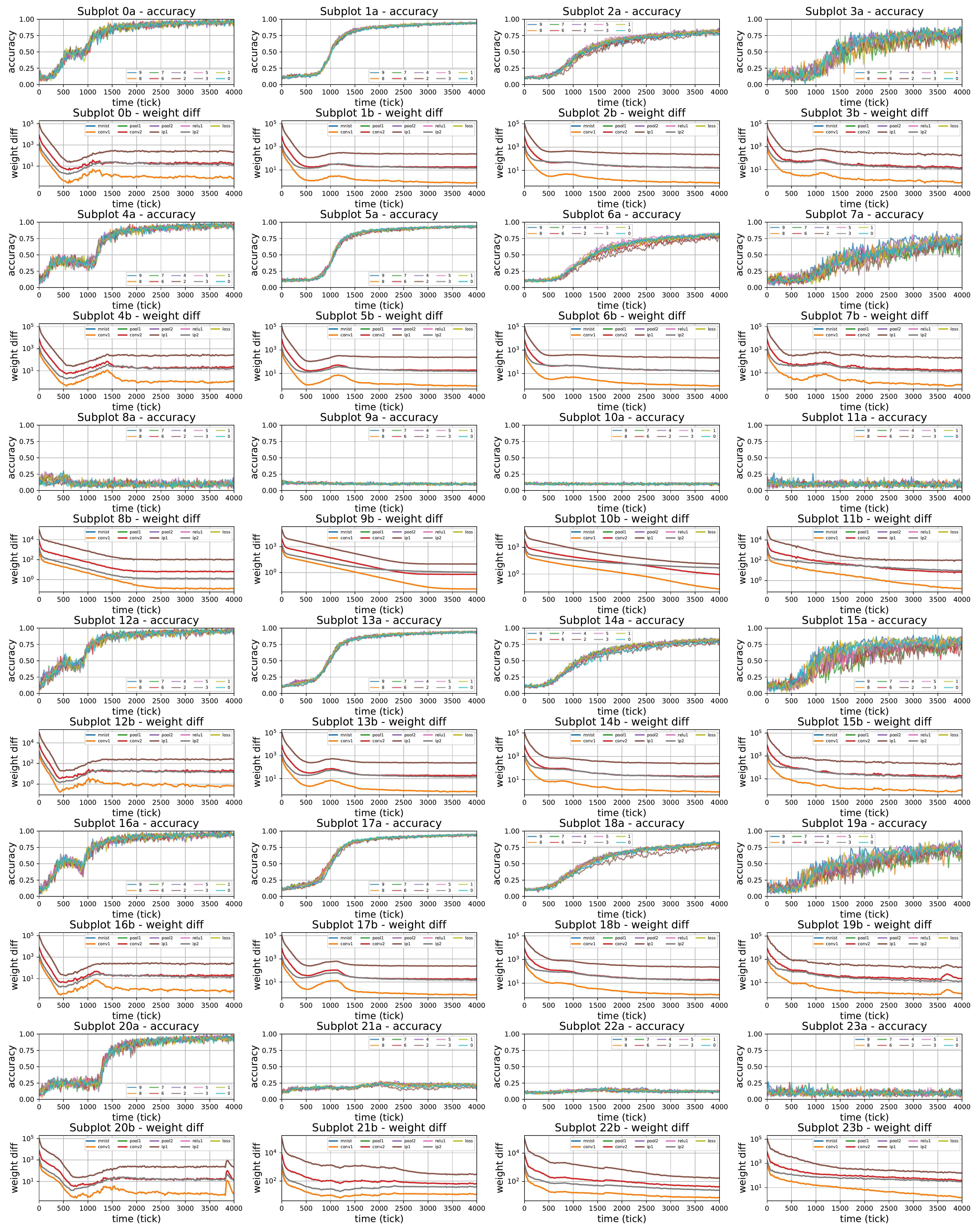}
    \caption{Simulation results for malicious and non-I.I.D. cases (vectorized image, zoom-in for details)}
    \label{fig:simulation_whole}
\end{figure}

Because the DFL network is a new computing architecture and we observed many unexpected behaviors and phenomena that never exist in normal federated ML, we present the accuracy and weight difference results from all combinations of non-I.I.D. cases and malicious cases \footnote{the configuration file of each combination is available in \href{https://github.com/twoentartian/DFL/tree/main/paper_result/non-iid-malicious-10nodes}{https://github.com/twoentartian/DFL/tree/main/paper\_result/non-iid-malicious-10nodes}. Notice that for non-I.I.D. cases described by the Dirichlet distribution, there are 10 simulation output folders, and each of them contains a configuration file and a plot of accuracy and weight difference. The subplot in Figure \ref{fig:simulation_whole} in this paper is the average of all plots in the corresponding output folders.} in Figure \ref{fig:simulation_whole}. Table \ref{tab:whole_plot_location} summarizes all combinations of non-I.I.D. and malicious cases and corresponding subplot locations in Figure \ref{fig:simulation_whole}. For example, subplot 13 is located at column 2 and row 4 in Table \ref{tab:whole_plot_location}, "(column 2, row 4)" also represents the subplot location in Figure \ref{fig:simulation_whole}. Each subplot has two parts: part a is the accuracy graph, and part b is the model difference among all nodes.

\begin{table}[h]
\centering
\caption{The subplot location for Figure \ref{fig:simulation_whole}}
\label{tab:whole_plot_location}
\begin{tabular}{|c|c|c|c|c|}
\hline
 & IID & \begin{tabular}[c]{@{}l@{}}non-I.I.D.\\ alpha=1\\ (10 repetitions) \end{tabular} & \begin{tabular}[c]{@{}l@{}}non-I.I.D.\\ alpha=0.1\\ (10 repetitions) \end{tabular} & \begin{tabular}[c]{@{}l@{}}non-I.I.D.\\ 1node2labels\end{tabular} \\ \hline
\begin{tabular}[c]{@{}l@{}}no malicious\\ HalfFedAvg\end{tabular} & Subplot 0 & Subplot 1 & Subplot 2 & Subplot 3 \\ \hline
\begin{tabular}[c]{@{}l@{}}dataset poisoning\\ HalfFedAvg\end{tabular} & Subplot 4 & Subplot 5 & Subplot 6 & Subplot 7 \\ \hline
\begin{tabular}[c]{@{}l@{}}model poisoning\\ HalfFedAvg\end{tabular} & Subplot 8 & Subplot 9 & Subplot 10 & Subplot 11 \\ \hline
\begin{tabular}[c]{@{}l@{}}no malicious\\ Reputation-0.05\end{tabular} & Subplot 12 & Subplot 13 & Subplot 14 & Subplot 15 \\ \hline
\begin{tabular}[c]{@{}l@{}}dataset poisoning\\ Reputation-0.05\end{tabular} & Subplot 16 & Subplot 17 & Subplot 18 & Subplot 19 \\ \hline
\begin{tabular}[c]{@{}l@{}}model poisoning\\ Reputation-0.05\end{tabular} & Subplot 20 & Subplot 21 & Subplot 22 & Subplot 23 \\ \hline
\end{tabular}%
\end{table}

Figure \ref{fig:simulation_whole} shows that the accuracy of ML models can reach over 0.9 under non-I.I.D. and dataset poisoning cases after 4000 ticks. The "epoch" concept is not suitable for DFL because all training samples are randomly selected from the training dataset and injected to the DFL nodes. Randomly choosing training data is closer to practice because each node has no information about the dataset of other nodes in a decentralized network. The total data samples trained by the DFL network (10 DFL nodes) is $ \frac{4000 \cdot 64 \cdot 10}{10} = 256000 $ at 4000 ticks, approximately 4.2 epoch for MNIST training dataset. The accuracy variances in the second and third columns seem to be smaller than those in the fourth column because the accuracy plot in the second and third columns is the averaged result from ten repetitions.

Subplot 8 and Subplot 20 show that \texttt{Reputation-0.05} mechanism can prevent model poisoning in I.I.D. cases, but it does not work in non-I.I.D. cases as shown in Subplots 21,22,23. The factors for successfully training models in non-I.I.D. and model poisoning situations include reputation algorithm, network topology, and the dataset distribution of each node. We have not dived too deep into non-I.I.D. and malicious cases because the reputation mechanism is not our focus. The DFL framework provides a reputation SDK to facilitate further research on reputation algorithms. 

There are some other phenomena in these subplots, which we will briefly discuss in this paper. These phenomena can lead to future research topics:

\begin{itemize}
    \item There is only one model in traditional ML systems and federated ML systems, indicating the model weight difference should be zero when a model is sufficiently trained. However, the model weight difference in DFL does not converge to zero but rather converges to a constant value when the model is properly trained. Subplots 8,9,10,11 also show unsuccessful training has lower model weight difference than successful training. In addition, the model weight difference does not constantly decrease during the training process. The weight difference at 4000 tick is even higher than 600 tick in Subplot 0. In other words, the training result of the DFL network is a set of models rather than a single model, and each model has similar accuracy. One of the potential applications of this property is training models whose dataset relates to geographical location. For example, a disease in different continents has different disease behavior, and there are two DFL nodes on each continent. Both DFL nodes should be able to train a model that can best predict the disease based on different disease behavior. For traditional ML systems, the geographical information must be included in the training dataset, but we would not know that geographical information is critical to the model if we do not aggregate all the data.
    
    \item There is a complex relationship between model weight difference and model accuracy, which we are not yet able to describe yet. The complexity is visible in Subplots 15 and 19 which show that the dataset distribution can change the trend of the model weight difference.
    
    \item In Subplots 0, 4, 12, and 16, there are regions where the accuracy plateaus at 0.5 from 500 ticks to 1000 ticks. The length of the plateau regions scales up with the number of nodes\footnote{this result is based on simulations not shown in this paper}, and it can greatly slow down the ML training when scaling to much larger DFL networks, which requires further investigation. 
    These plateau regions are one of the reasons we avoid simulating large-scale DFL networks because they will dominate the training performance. 
\end{itemize}

\subsection{The speed ratio of model training and model updating}
This section investigates the trade-off between bandwidth requirements and ML performance. This is done using simulations to explore the impact of the ratio of model training interval over model updating interval on ML performance. The interval of model training is controlled by training batch size and data injection rate. For example, injecting 64 training samples per 10 ticks with a training batch size of 64 results in one training iteration per 10 ticks. The interval of model updating is controlled by FedAvg buffer size, the number of peers, and the training interval of peers. 

A new node type called "observer" is used in this simulation. The observer node does not train ML models and only performs FedAvg on received models. The observer can ensure that there are enough communications among nodes and that the final model is a combination of training and updating rather than an independently trained model.

\begin{figure}[h]
    \centering
     \includegraphics[width=\textwidth]{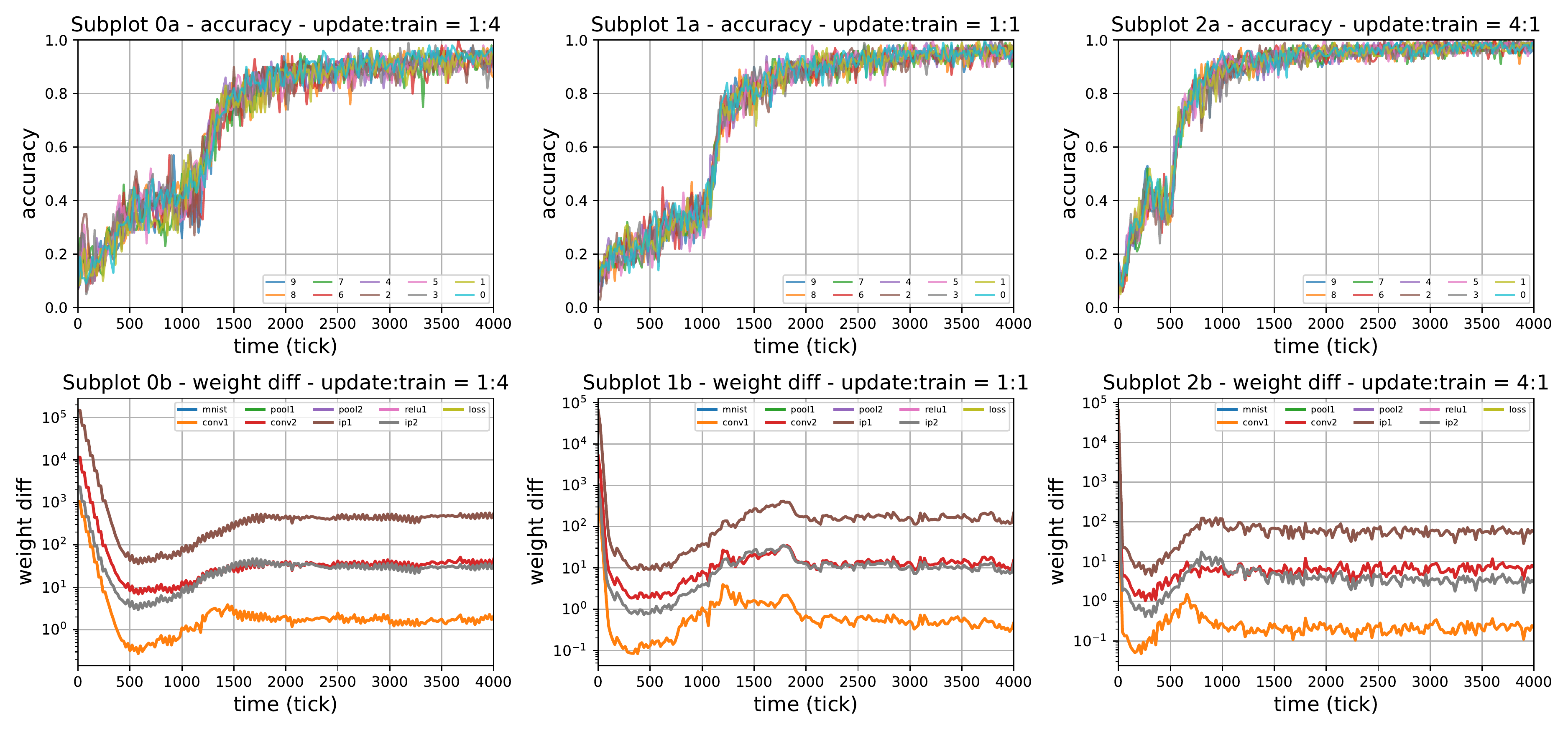}
    \caption{Update/train ratio simulation results (update/train = 1:4,1:1,4:1)}
    \label{fig:simulation_update_train}
\end{figure}

Some modifications \footnote{the configuration files are available in \href{https://github.com/twoentartian/DFL/tree/main/paper_result/update_train_ratio}{https://github.com/twoentartian/DFL/tree/main/paper\_result/update\_train\_ratio}} are made to the configuration in Section \ref{subsection:ml_propertires}: 1) simulations use a fully-connected 10-node DFL network; 2) each node uses an I.I.D. dataset; 3) the first node is configured as an observer; 4) the reputation algorithm is "\texttt{Reputation-0.05}". Because the training interval is kept as 8 to 12 ticks, and each node should get 8 models in approximately 10 ticks, we set the FedAvg buffer size to 32, 8, and 2 to achieve different update/train ratios (1:4, 1:1, and 4:1). The result is shown in Figure \ref{fig:simulation_update_train}.

\begin{figure}[h]
    \centering
     \includegraphics[width=\textwidth]{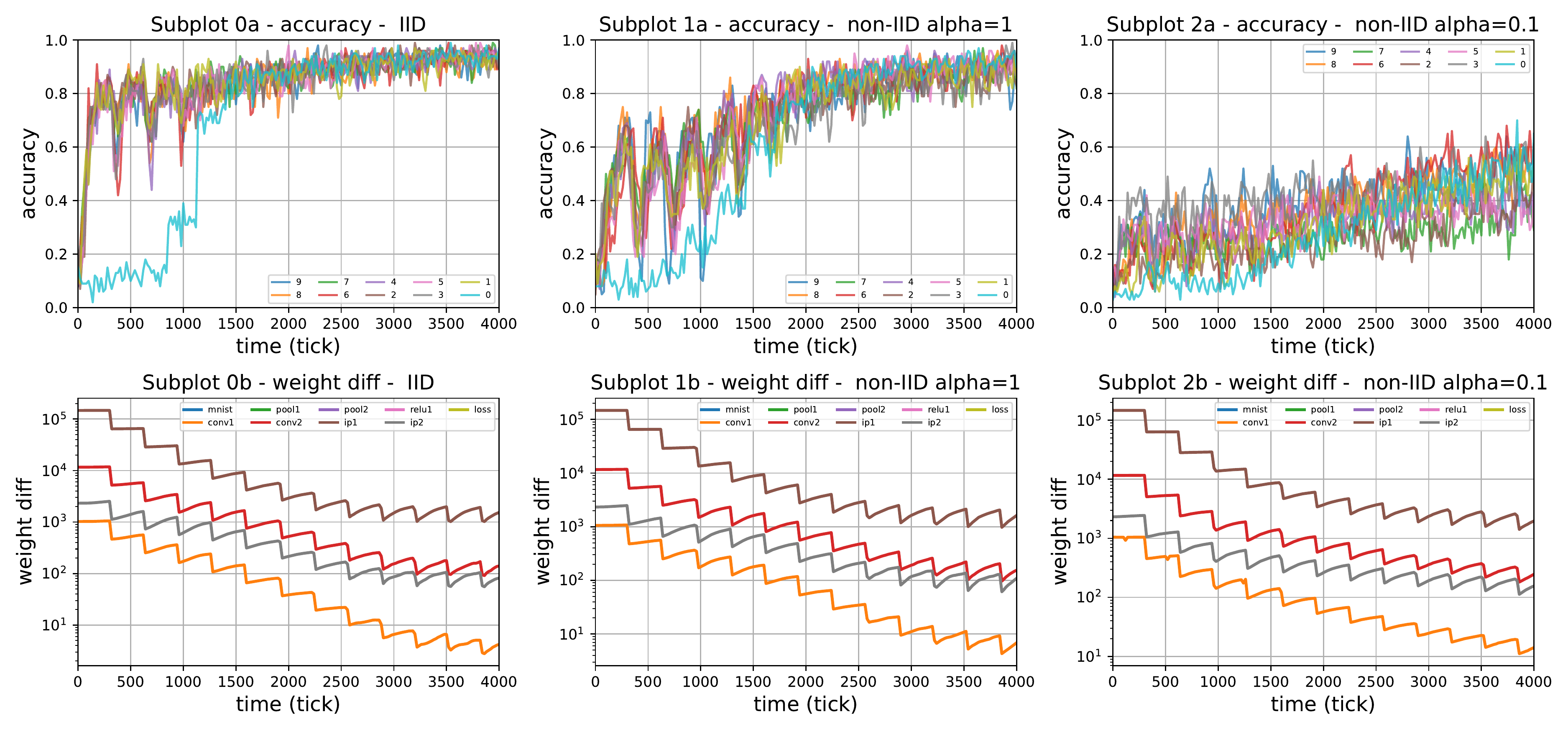}
    \caption{Update/train=1:32 simulation results (I.I.D., alpha=1, alpha=0.1)}
    \label{fig:simulation_update_train_1_32}
\end{figure}

Figure \ref{fig:simulation_update_train} indicates that more communication among nodes can accelerate the training. It is a trade-off between training speed and bandwidth because more communication requires more bandwidth consumption. We also test several extreme situations where the update/train ratio is 1:32 and the dataset distributions are I.I.D., alpha=1 and alpha=0.1, as shown in Figure \ref{fig:simulation_update_train_1_32}. The third subplot shows that the accuracy has already increased but needs more communication. Figure \ref{fig:simulation_update_train_1_32} shows it is possible to train a model with low-frequency communications, such as in embedded systems and mobile systems.

\section{conclusion}
\label{conclusion_future_work}

This paper presents a new asynchronous federated learning system, integrating a specialized blockchain system to generate the proof of contribution on the ML model. Using blockchain as a distributed proof system rather than a distributed ledger system removes the synchronicity requirement to increase performance. Meanwhile the asynchronicity also introduces the partial aggregation of models in training phase. We implement the architecture as DFL, which contains an executable prototype, a simulator to investigate ML behavior and an SDK to implement and test reputation algorithms. In a four node experiment with MNIST and LeNet, the training accuracy can be up to 90\% while the blockchain overhead can be kept limited to within only 5\% of total execution time. The simulation results show reasonable training accuracy under both non-I.I.D. datasets as well as malicious dataset attacks. However, the reputation mechanism only works in certain model attack cases and requires further research.

\bibliographystyle{ACM-Reference-Format}
\bibliography{reference}

\section*{Acknowledgment}

We want to thank Xun Gui and Zixuan Xie for their discussions and ideas on blockchain and machine learning. Their deep insights in blockchain and machine learning helped improve this project. We also thank Kewei Du, Xiaoning Shi and Li Xu for reviewing this paper and providing precious feedback. This research was performed with the support of the European Commission, the ECSEL JU project NewControl (grant no. 826653) and the Eureka PENTA project Vivaldy (grant no. PENT191004).

\end{document}